\documentclass[conference]{IEEEtran}
 \IEEEoverridecommandlockouts

\usepackage{cite}

\onecolumn

\usepackage{amsmath}
\usepackage{amsfonts}
\usepackage{amssymb}
\usepackage{amsthm}  
\usepackage{bbm}  
\usepackage{color}
 \usepackage{amssymb}
\usepackage{txfonts}
\usepackage[T1]{fontenc}

\newtheorem{thm}{Theorem}

\newtheorem{Lemma}{Lemma}

\newtheorem{Claim}{Claim}
\theoremstyle{remark}
\newtheorem{Remark}{Remark}

\newtheorem{Definition}{Definition}

\ifCLASSINFOpdf
   \usepackage[pdftex]{graphicx}
\else
\fi
\usepackage{makecell}
\usepackage{multirow}
\usepackage{tikz}
\newcolumntype{x}[1]{>{\centering\arraybackslash}p{#1}}

\begin{document}
%
\title{Trade-off between Communication and Cooperation in the Interference Channel}

\author{Farhad Shirani
  \IEEEauthorblockN{}
  \IEEEauthorblockA{EECS Department\\University of Michigan\\ Ann Arbor,USA \\
    Email: fshirani@umich.edu } 
 
  \and
  \IEEEauthorblockN{S. Sandeep Pradhan}
  \IEEEauthorblockA{EECS Department\\University of Michigan\\ Ann Arbor,USA \\
    Email: pradhanv@umich.edu}
}


%


\maketitle

\begin{abstract}
 We consider the problem of coding over the multi-user Interference Channel (IC). 
It is well-known that aligning the interfering signals results in improved achievable rates in 
certain setups involving more than two users.  We argue that in the general interference problem, 
senders face a tradeoff between communicating their message to their corresponding decoder or cooperating 
with other users by aligning their signals. Traditionally, interference alignment is carried out using 
structured codes such as linear codes and group codes. We show through an example that the usual structured coding 
schemes used for interference neutralization lack the necessary flexibility to optimize this tradeoff. 
Based on this intuition, we propose a new class of codes for this problem. We use the example to show 
that the application of these codes gives strict improvements in terms of achievable rates. Finally,  
we derive a new achievable region for the three user IC which strictly improves upon the previously known 
inner bounds for this problem. 

\end{abstract}

\section{Introduction}

\IEEEPARstart{T}{he} interference channel problem describes a setup where multiple 
pairs of transmitters and receivers share a communication
medium. Each receiver is only interested in decoding the message from its corresponding transmitter. However, 
since the channel is shared, signals from other senders interfere with the desired signal at each decoder. The
 presence of interfering signals adds new dimensions to this problem in terms of strategies that can be 
used as compared to point-to-point (PtP) communication.   
For example, the encoders can  \textit{cooperate} with each other  by choosing their channel inputs in a way that 
would facilitate their joint communication. It turns out that,  often, this cooperation requires 
an encoder to employ a strategy which  may be sub-optimal from its own PtP communications perspective. 
In this paper, we investigate this tradeoff and develop a new class of codes which allow for 
more efficient cooperation between the transmitters.

Characterizing the capacity region for the general IC has been a challenge for decades. Even in 
the simplest case of the two user IC, the capacity region is only known in special cases \cite{Sato}\cite{Costa}. 
The best known achievable region for the IC was due to Han and Kobayashi \cite{HK}. However, recently it was shown 
that the Han-Kobayashi (HK) rate region is suboptimal \cite{chandra}\cite{IC}.  Particularly, when there are more 
than two transmitter-receiver pairs, the natural generalization of the HK strategy can be improved upon by inducing 
structure in the codebooks used in the scheme \cite{IC}. Structured codes such as linear codes and group codes enable 
the encoders to align their signals more efficiently. This in turn reduces interference at the decoders. Such codebook 
structures have also proven to give gains in other multi-terminal communication problems \cite{KM}-\cite{MD}.   

\begin{figure}[!t]
\centering
\includegraphics[height=1.3in]{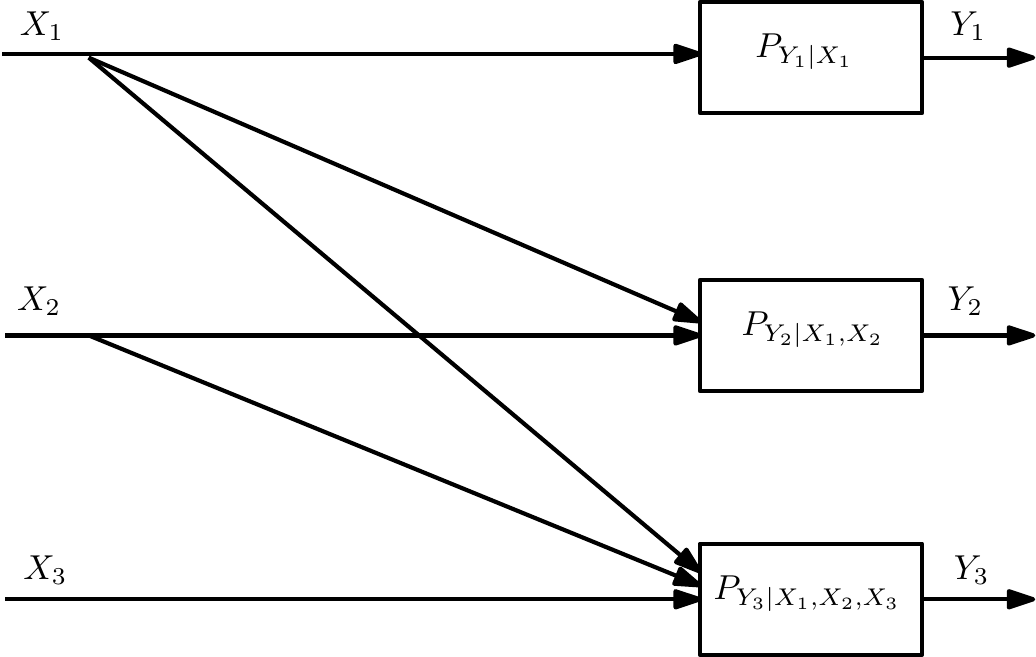}
\caption{A setup where interference alignment is beneficial to user 3 but harmful for user 2.}
\label{fig:1}
\end{figure}

The idea of interference alignment was proposed for managing interference when there are three or more users.
 Initially, the technique was proposed by Maddah-Ali et. al. \cite{MA} for the MIMO X channel, and for the 
multi-user IC by Jafar and Cadambe \cite{CJ}. The interference alignment strategy was developed for cases of 
additive interference and uniform channel inputs over finite fields. This was extended to arbitrary interference 
settings and input distributions in \cite{IC}. However, it turns out that alignment is not always beneficial 
to the users in terms of achievable rates. Consider the example in Figure \ref{fig:1}. Intuitively, it 
would be beneficial to align the input from users 1 and 2 to reduce interference at decoder 3. However, 
if users 1 and 2 align their signals, it becomes harder for decoder 2 to distinguish between the two 
inputs. One might suggest that the problem could be alleviated if users 1 and 2 designed their codebooks 
in a way that they would "look" aligned at decoder 3 based on $P_{Y_3|X_1,X_2,X_3}$, but at the same time they 
would seem different at decoder 2 based on $P_{Y_2|X_1,X_2}$. In this paper we show that linear codes lack the 
necessary flexibility for such a strategy. Based on this intuition, we propose a new class of structured codes. 
Using these codes we derive an achievable rate region which improves upon the best known achievable region for 
the three user IC given in \cite{IC}.

The rest of the paper is organized as follows: Section \ref{sec:not} gives the notation used in the paper as well as the problem statement. In section \ref{sec:trade}, we consider two different examples of three user IC. In the first example - where interference alignment is strictly beneficial - we prove that not only structured codes are useful for alignment but that any arbitrary coding scheme which achieves optimality must possess certain linearity properties. In the second example, we show the existence of the tradeoff discussed above and prove that linear codes are suboptimal for that example. Section \ref{sec:code}, gives the new codebook constructions and proves that these new codes outperform the linear coding scheme in \cite{IC}. In Section \ref{sec:RR}, we provide a new general achievable rate region for this problem. Section \ref{sec:conclude}, concludes the paper.

\section{Problem Statement and Notation}
\label{sec:not}
In this section, we give the notation used in the paper and provide the problem statement. Throughout the paper, we denote random variables by capital letters such as $X, U$, their realizations by small letters $x,u$, and their corresponding alphabets (finite) by sans-serif typeface $\mathsf{X}$, $\mathsf{U}$, respectively. Small letters such as $l, k$ are used to represent numbers. The field of size $q$ is denoted by $\mathbb{F}_q$. We represent the field addition by $\oplus$ and the addition on real numbers by $+$.
For $m\in \mathbb{N}$, We define the set of numbers $[1,m]\triangleq \{1,2,\dotsb, m\}$. Vectors are represented by the bold type-face such as $\mathbf{u}, \mathbf{b}$. For a random variable $X$, $A_{\epsilon}^n(X)$ denotes the set of $\epsilon$-typical sequence of length $n$ with respect to the probability distribution $P_X$, where we use the definition of frequency typicality. Let $q$ be a prime number. For $l\in\mathbb{N}$, consider $U_i,i\in[1,m]$ i.i.d random variables with distribution $P_U$ defined on the field $\mathbb{F}_q$. $U^{\otimes{l}}$ denotes a random variable which has the same distribution as $\sum_{i\in [1,l]}U_i$ where the summation is over $\mathbb{F}_q$. 

We proceed with formally defining the three user IC problem.
A three user IC consists of three input alphabets $\mathsf{X}_i, i\in \{1,2,3\}$, three output alphabets $\mathsf{Y}_i, i\in \{1,2,3\}$, and a transition probability matrix $P_{\mathbf{Y}|\mathbf{X}}$. A code for this setup is defined as follows.

\begin{Definition}
 A three user IC code $(n,\mathsf{M}_1,\mathsf{M}_2,\mathsf{M}_3, \mathbf{e},\mathbf{d})$ consists of (1) Three sets of message indices $\mathsf{M}_i$ (2) Three encoder mappings $e_i:\mathsf{M}_i\to \mathsf{X}_i^n,i\in[1,3]$, without loss of generality, these maps are assumed to be injective (3) and three decoding functions $d_i:\mathsf{Y}_i^n\to \mathsf{M}_i, i\in [1,3]$. We define the codebook corresponding to the encoding map $e_i$ as $\mathbb{C}_i=\{e_i(m_i)|m_i\in \mathsf{M}_i\},i\in[1,3]$. The rate of user $i$ is defined as $r_i=\frac{1}{n}\log{|\mathbb{C}_i|}$.
\end{Definition}
\begin{Definition}
 A rate-triple $(R_1,R_2,R_3)$ is said to be achievable if for every $\epsilon>0$, there exists a code $(n,\mathsf{M}_1,\mathsf{M}_2,\mathsf{M}_3, \mathbf{e},\mathbf{d})$ such that $(1)$ $r_i\geq R_i-\epsilon, i \in [1,3]$, and (2) $P(d(\mathbf{Y}^n)=\mathbf{M}|\mathbf{e}(\mathbf{M})=\mathbf{X}^n)\geq 1-\epsilon$.
\end{Definition}

 We make frequent use of coset codes and Nested Linear Codes (NLC) which are defined next.
\begin{Definition} 
 A  $(k,n)$ coset code $\mathcal{C}$ is characterized by a generator matrix $G_{k\times n}$ and a dither $\mathbf{b}^n$ defined on the field $\mathbb{F}_q$. The code is defined as $\mathcal{C}\triangleq\{\mathbf{u}G\oplus \mathbf{b}|\mathbf{u}\in \mathbb{F}_q^k\}$. 
The rate of the code is given by $R=\frac{k}{n}\log{q}$.
\end{Definition}
\begin{Definition}
\label{def:PNLC}
For natural numbers $k_i<k_o,k'_o<n$, let $G_{k_i\times n}, \Delta{G}_{(k_o-k_i)\times n}$ and $\Delta{G'}_{(k'_o-k_i)\times n}$ be matrices on $\mathbb{F}_q$. Define $\mathcal{C}_i, \mathcal{C}_o$ and $\mathcal{C}'_o$ as the linear codes generated by $G$, $[G|\Delta{G}]$ and $[G|\Delta{G}']$, respectively.  $\mathcal{C}_o$ and $\mathcal{C}'_o$ are called a pair of NLC's with inner code $\mathcal{C}_i$. We denote the outer rates as $r_o=\frac{k_o}{n}$ and $r'_o=\frac{k'_o}{n}$ and the inner rate $r_i=\frac{k_i}{n}$.
  \end{Definition}
\section{The Interference Alignment tradeoff}
\label{sec:trade}
In this section, we investigate the interference alignment tradeoff mentioned in the introduction in more detail. We show that in certain three user interference setups, on the one hand,  alignment is beneficial to one of the users, while on the other hand, the rates achieved by the aligning users is reduced due to the alignment. We investigate the phenomenon in two examples. The first example involves a three user interference setup. In this example, the first two encoders use linear codes to manage the interference for the third user. This gives a strictly improved achievable rate region. It is well-known that interference alignment can be induced efficiently by the application of structured codes. Additional to this, we show the stronger statement that the only ensemble of codes which achieve the desired rate-triples in this example, are the ones with specific linearity properties. 
Next, we build upon the first example to create a setup where alignment is beneficial to one of the users and harmful for the other one. This second example provides the motivation for our new codebook constructions in the next section.
\subsection{Example 1}
Consider the example shown in Figure \ref{fig:goodalignment}. All of the inputs are $q$-ary and the additions are defined on the field $\mathbb{F}_q$. The three outputs of the channel are $Y_i=X_i\oplus  N_1\oplus  N_3, i\in\{1,2\}$, $Y_3=X_1\oplus  X_2\oplus  X_3\oplus  N_3$. 
We are interested in achieving the following rates for the first and second users:
\begin{figure}[!h]
\centering
\includegraphics[height=1.3in]{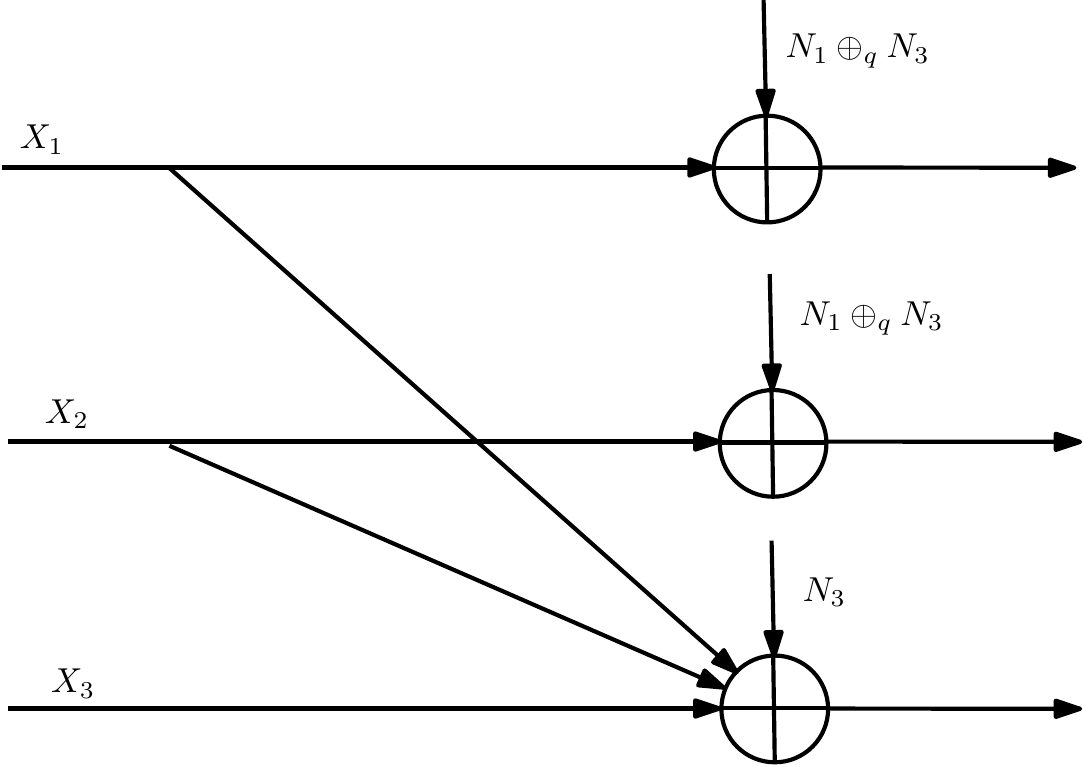}
\caption{A Three User IC Where Alignment Is Strictly Beneficial}
\label{fig:goodalignment}
\end{figure}
\vspace{-0.1in}
\begin{equation*}
 R_1=R_2=\log{q}-H(N_1\oplus N_3).
\end{equation*}
Given these rates, we want to maximize $R_3$. The following lemma shows that linear codes achieve the optimum $R_3$ for this setup. Furthermore, we show that if an ensemble of codes achieves the optimum $R_3$, then the codes corresponding to the first two users  are "almost" the same coset code.
\begin{Lemma}
\label{Lem:eg1}
 For a given family of codes $(n,\mathsf{M}_1,\mathsf{M}_2,\mathsf{M}_3, \mathbf{e},\mathbf{d}), n \in \mathbb{N}$ satisfying the rate and error constraints at decoders 1 and 2, user $3$ can achieve the rate $R_3=H(N_1\oplus N_3)-H(N_3)$ iff there exists a dither $\mathbf{b}$ such that for every random variable $N$ defined on $\mathbb{F}_q$ with positive entropy, the following holds:
\begin{equation}
 P(e_1(M_1)\oplus  e_2(M_2))\in \mathbb{C}_1\bigcup \mathbb{C}_2\oplus A_\epsilon^n(N)\oplus \mathbf{b})\to 1, \text{ as } n\to \infty.
\end{equation}
 
 Equivalently, the optimal rate is achieved iff there exists another family of codes $(n,\mathsf{M}'_1,\mathsf{M}'_2,\mathsf{M}'_3, \mathbf{e}',\mathbf{d}'), n \in \mathbb{N}$ for which
 1) $P(\mathbf{e'}_i(M'_i)\in \mathbb{C}_i\oplus A_\epsilon^n(N))\to 1$ as $n\to \infty$, 2) $\mathbb{C}'_1=\mathbb{C}'_2$ is a coset code, and 3) they also achieve the rate triple $(R_1,R_2,R_3)=(\log{q}-H(N_1\oplus  N_3),\log{q}-H(N_1\oplus  N_3),H(N_1\oplus N_3)-H(N_3))$. 
 \end{Lemma}
\begin{IEEEproof}
 We provide an outline of the proof in Appendix \ref{App:eg1}. 
\end{IEEEproof}

 The lemma proves that even if we expand our search to arbitrary $n$-length codebook constructions (as opposed to the usual random codebook generation based on single-letter distributions), coset codes are the only efficient ensemble of codes for the classes of interference channels under consideration up to small perturbations. This is a stronger assertion than the well-known result that linear codes are useful for aligning the interfering signals. The lemma can be used to provide a converse result proving that schemes involving random unstructured codes (e.g. the generalized version of the single-letter HK scheme), can't achieve the desired rate-triple without directly analyzing the bounds corresponding to their achievable rate region as done in \cite{IC}.

   \subsection{Example 2}
\begin{figure}[!h]
\centering
\includegraphics[height=1.3in]{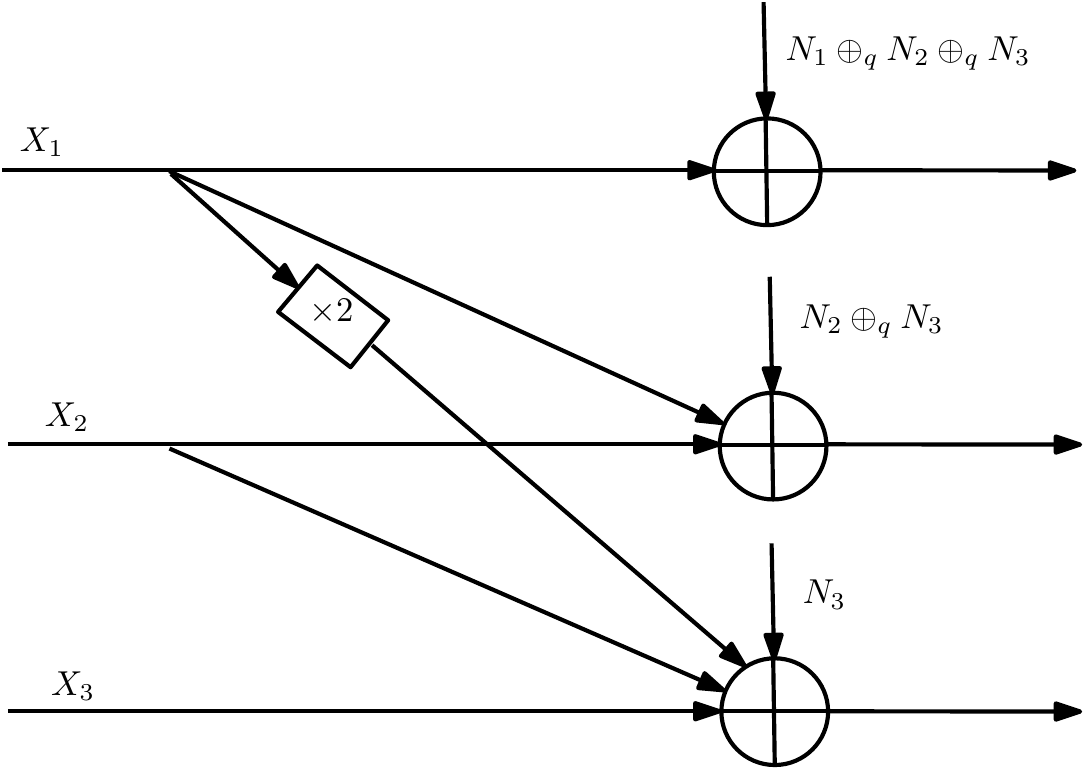}
\caption{A Three User IC Where Alignment Results in a Tradeoff}
\label{fig:eg2}
\end{figure}

Next, we consider an example where interference alignment results in a tradeoff between two of the users. Consider the setup in Figure \ref{fig:eg2}. Similar to the previous example, all input alphabets, output alphabets, and additions are defined on the field $\mathbb{F}_q$. The outputs of the channel are $Y_1=X_1\oplus_{q}N_1\oplus_{q}N_2\oplus_{q}N_3$, $Y_2=X_1\oplus_{q}X_2\oplus_{q}N_2\oplus  N_3$, and $Y_3=2X_1\oplus_{q}X_2\oplus_{q}X_3\oplus_{q}N_3$. Following our arguments in the previous example, for user 3 to be able to transmit its messages at rate $R_3=H(X_3\oplus  N_3)-H(N_3)$, the inputs for users $1$ and $2$ must align. However, if these two users align their inputs, user 2 would not be able to decode its message which is being corrupted by its aligned interfering signal coming from user 1. Hence, we have a tradeoff.  We proceed with evaluating the rate-triples achievable in this example. The following lemma proves that we must have $R_1+R_2\leq \log{q}-H(N_2\oplus  N_3)$, otherwise the rate-triple $(R_1,R_2,R_3)$ is not achievable.
\begin{Lemma}
\label{Lem:lem2}
Given that $(R_1,R_2,R_3)$ is achievable, we must have $R_1+R_2\leq log{q}-H(N_2\oplus  N_3)$. 
\end{Lemma}
\begin{IEEEproof}
 Since the rate triple is achievable, there exists a family of codes $(n,\mathsf{M}_1,\mathsf{M}_2,\mathsf{M}_3, \mathbf{e},\mathbf{d})$ for which the codewords sent by the second user, $X^n_2$, is decoded at decoder 2 with error probability approaching 0. Assuming errorless decoding at decoder 2, the decoder has access to $X^n_2, X_1^n\oplus X_2^n\oplus  N_2^n$. The decoder can subtract $X_2^n$ from $X_1^n\oplus X_2^n\oplus  N_2^n$ to get $X_1^n\oplus N_2^n$. Now, since by assumption decoder 1 can decode $X_1^n$ from $X_1^n\oplus N_1^n\oplus N_2^n$ with error going to 0, decoder 2 can use $X_1^n\oplus N_2^n$ to recover $X_1^n$ with small error. So decoder 2 has access to $M_1$ and $M_2$. By the converse of the point-to-point channel coding theorem, we must have $\frac{1}{n}\log{|\mathsf{M}_1\times\mathsf{M}_2|}\leq \log{q}-H(N_2)$, which completes the proof.
\end{IEEEproof}

 We want to achieve the rate $R_1=\log{q}-H(N_1\oplus N_2\oplus N_3)$. In other words, encoder 1 is to operate at PtP optimality. The goal is to optimize the linear combination $R_2+R_3$. We argue that the linear coding scheme presented in \cite{IC} can't achieve the triple $(R_1,R_2,R_3)$ for $R_2+R_3>H(N_1\oplus N_2\oplus N_3)-H(N_3)$. 

\begin{Lemma}
\label{Lem:lem3}
 Given $R_1=\log{q}-H(N_1)$, the scheme in \cite{IC} can't achieve $R_2+R_3>H(N_1\oplus N_2\oplus N_3)-H(N_3)$.  
\end{Lemma}
\begin{IEEEproof}
 We provide the intuition behind the proof here. Let us use two NCL's $\mathcal{C}_1$ and $\mathcal{C}_2$ as defined in Definition \ref{def:PNLC} to transmit the messages at encoders 1 and 2. Let the rate of $\mathcal{C}_j, j\in \{1,2\}$ be $r_j$ and let the inner code have rate $r_i$. If we assume that the coding scheme exists which achieves the rate-triple, then by the proof of Lemma \ref{Lem:lem2}, one should be able to recover $X_1, X_2$ from $X_1\oplus X_2\oplus N_2\oplus N_3$  with small error probability. Also, at decoder $3$, the decoder can reconstruct $X_3$ with low error probability and by subtraction it can have $2X_1\oplus X_2\oplus N_3$. Note that in the linear coding scheme, both $2X_1\oplus X_2$ and $X_1\oplus X_2$ come from randomly and uniformly generated linear codes of rate $r_1+r_2-r_i$. So, given that $X_1$ and $X_2$ can be recovered from $X_1\oplus X_2\oplus N_2\oplus N_3$, decoder 3 must be able to recover $X_1$ and $X_2$ from $2X_1\oplus X_2\oplus N_3$. Then similar to the proof of Lemma \ref{Lem:eg1}, by the point-to-point channel coding converse, we must have $R_1+R_2+R_3<\log{q}-H(N_3)$.
\end{IEEEproof}

The arguments in the proof of the previous lemma suggest that NLC's lack the necessary flexibility when it comes to determining the size of different linear combinations of such codes. We explain this in more detail. Consider two NLC's, $\mathcal{C}$ and $\mathcal{C}'$, with rates $r_o$ and $r'_o$, respectively, and with inner code rate $r_i$. The rate of any linear combination of the two, $\alpha \mathcal{C}\oplus  \beta\mathcal{C}', \alpha,\beta\in\mathbb{F}_q\backslash\{0\}$, is equal to $r_o+r'_o-r_i$. Whereas in settings such as the one at hand, it is desirable to have different rates for different values of $\alpha$ and $\beta$. In this setup, decoder 2 requires $\mathcal{C}_1\oplus  \mathcal{C}_2$ to be large (since by Lemma \ref{Lem:lem2} in order to increase $R_2$ it needs to increase the rate of this linear combination) and decoder 3 wants the size of the interfering codebook $2\mathcal{C}_1\oplus  \mathcal{C}_2$ to be small, so that it can decode the interference. 
 In the next section, we provide a new class of codes. The new construction allows for different rates for different linear combinations of such codes. This in turn results in higher achievable sum-rates.
 \section{A New Class of Code Constructions}
 \label{sec:code}
In this section, we present our new coding constructions. These new codes are called Quasi Linear Codes (QLC). They are not linearly closed but maintain a degree of linearity. In order to construct a QLC, we first construct a linear code. Then, we take a subset of that codebook to transmit the messages. More precisely, QLC's are defined as follows: 
\begin{Definition} 
\label{def:Qlingen}
A $(k,n)$ QLC on the field $\mathbb{F}_q$, is characterized by a generator matrix $G_{k\times n}$, a dither $\mathbf{b}^n$ and a set $\mathsf{U}$. The codebook is defined as 
\begin{equation*}
 \mathcal{C}\triangleq\{\mathbf{u}G\oplus \mathbf{b}|\mathbf{u}\in \mathsf{U}\}.
 \end{equation*}
  For injective $G$ on $\mathsf{U}$, the rate of the code is given by $R=\frac{1}{n}\log{|\mathcal{C}|}=\frac{1}{n}\log{|\mathsf{U}|}$.
\end{Definition}
In this paper we only consider the cases when $\mathsf{U}$ is a cartesian product of typical sets: 
\begin{equation*}
\mathcal{C}\triangleq \{\sum_{i\in [1,m]}\mathbf{u}_iG_i\oplus \mathbf{b}|\mathbf{u}_i\in A_{\epsilon}^{k_i}(U_i)\}. 
\end{equation*}
A pair of Nested Quasi Linear Codes (NQLC) is defined below:
\begin{Definition}
\label{Def:PQLC}
For natural numbers $k_1,k_2,\dotsb, k_m$, let $G_{k_i\times n}, i\in [1,m]$ be matrices, and $\mathbf{b}, \mathbf{b}'$ dithers all defined on $\mathbb{F}_q$ . Also, let $(U_1,U_2,\dotsb, U_m)$ and $(U'_1,U'_2,\dotsb, U'_m)$ be a pair of random vectors on $\mathbb{F}_q$. The pair of QLC's characterized by the matrices $G_{k_i\times n}, i\in [1,m]$ and each of the two dithers and vectors of random variables are called a pair of NQLC's.
  \end{Definition}
As explained in the previous section, our motivation for defining NQLC's, is to construct codes such that different linear combinations of those codes have different rates. The next lemma shows that NQLC's have this property.
 \begin{Lemma}
 Let $\mathcal{C}$ and $\mathcal{C}'$ be two QLC's as defined in Definition \ref{Def:PQLC}, whose generator matrices and dithers are taken randomly and uniformly from $\mathbb{F}_q$. Then, $\alpha\mathcal{C}_1\oplus \beta\mathcal{C}_2$ has rate close to $\sum_{i\in [1,m]}\frac{k_i}{n}H(\alpha U_i\oplus \beta U'_i)$ for large $n$ with high probability.
\end{Lemma}
\begin{IEEEproof}
 The proof follows from the injectiveness of the $G_i$'s and the usual typicality arguments and is omitted.
\end{IEEEproof}
Having defined QLC's, we return to our interference channel setup in Example 2. We claim that NQLC's can achieve a sum-rate $R_2+R_3$ which is higher than $H(N_1\oplus N_2\oplus N_3)-H(N_3)$. 
\begin{Lemma}
\label{Lem:newrate}
There exists achievable rate-triples $(R_1,R_2,R_3)=(\log{q}-H(N_1\oplus N_2\oplus N_3), r_2, r_3)$ such that $r_2+r_3>H(N_1\oplus N_2\oplus N_3)-H(N_3)$.
\end{Lemma}
\begin{IEEEproof}
Refer to Appendix \ref{App:newrate}.
 
\end{IEEEproof}
So far we have proved that NQLC's outperform NLC's in this specific example. It is straightforward to show that NQLC's are a generalization of NLC's. To see this, consider an arbitrary pair of NLC's with the parameters as in definition \ref{def:PNLC}. These two codes are a pair of NQLC's with parameters $m=3$, $U_1, U_2$ and $U'_1,U'_3$ uniform, $U_3$ and $U'_2$  constants and $k_1=k'_1=k_i$ and $k_2=k_o-k_i, k'_3=k'_o-k_i$. So, any rate region achievable by NLC's is also achievable using NQLC's. 
\section{New Achievable Rate Region}
\label{sec:RR}
\begin{figure}[!h]
\centering
\includegraphics[height=1in]{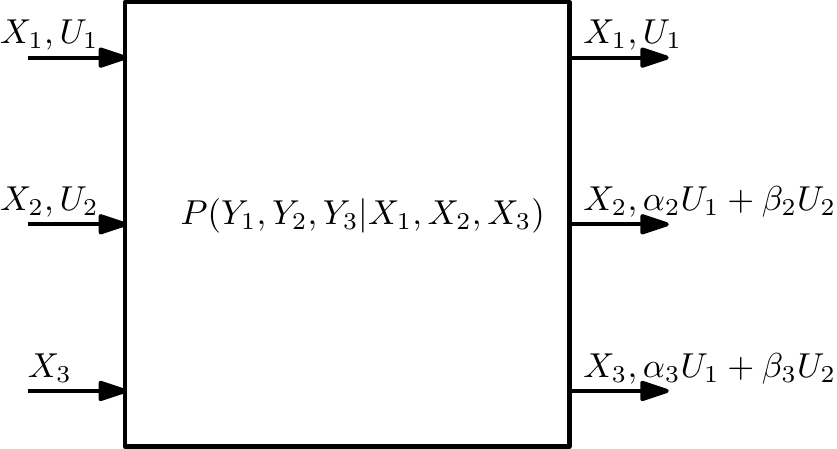}
\caption{The LHS random variables are the ones sent by each encoder, the RHS random variables are the ones decoded at each decoder.}
\label{fig:rvs}
\end{figure}

In this section, we provide a general achievable rate region for the three user IC. The scheme is similar to the one presented in \cite{IC} (Theorem 2). The main difference is that here instead of NLC's we use NQLC's. The random variables involved in the coding scheme are depicted in Figure \ref{fig:rvs}. Note that in contrast with the scheme in \cite{IC}, decoder 2 reconstructs a linear combination of $U_1$ and $U_2$. By setting $\alpha_2=0, \beta_2=1$, we recover the random variables in \cite{IC}. The next theorem provides the achievable rate region. 

\begin{Definition}
 For a given three user IC problem with $q$-ary inputs and outputs, define the set $\mathcal{R}_{3\text{-IC}}$ as the set of rate triples $(R_1,R_2,R_3)$ such that there exist 1) a joint probability distribution $P_{U_1,X_1}P_{U_2,X_3}P_{X_3}$, 2) A vector of positive reals $(K_1,K_2,L_1,L_2,T_1,T_2)$, and 3) a vector of parameters $(m,n,k_1,k_2,\dotsb,k_m)$ and pair of vectors of random variables $(V_{ij})_ {j\in [1,m]}, i\in\{1,2\}$, such that the following inequalities are satisfied:
\begin{align}
 &R_1=L_1+T_1, R_2=L_2+\frac{I(U_2;\alpha_2U_1\oplus \alpha_2U_2)}{H(U_2)}T_2\label{eq:rates}\\
 &r_{1,0}-T_1\geq \log{q}-H(U_1), r_{0,1}-T_2\geq \log{q}-H(U_2), \label{eq:packS}\\
 &K_1+r_{1,0}-T_1\geq \log{q}+H(X_1)-H(X_1,U_1)\\
 &K_2+r_{0,1}-T_2\geq \log{q}+H(X_2)-H(X_2,U_2)\label{eq:packF}\\
 &r_{0,1}\leq \log{q}-H(U_1|X_1,Y_1)\label{eq:cov1S}\\
 &r_{0,1}+L_1+K_1\leq \log{q}+H(X_1)-H(U_1,X_1|Y_1)\\
 &L_1+K_1\leq I(X_1;U_1Y_1)\label{eq:cov1F}\\
 &r_{\alpha_2,\beta_2}\leq \log{q}-H(\alpha_2U_1\oplus \beta_2|X_2,Y_2)\label{eq:cov2S}\\
 &r_{\alpha_2,\beta_2}\!\!+L_2+K_2\leq \log{q}\!+\!H(X_2)-H(\alpha_2U_1\!\!\oplus \!\!\beta_2U_2,X_2|Y_2)\\
 &L_2+K_2\leq I(X_2;U_2Y_2)\label{eq:cov2F}\\
 &r_{\alpha_3,\beta_3}\!\!+L_3+K_3\leq \log{q}\!+\!H(X_3)-H(\alpha_3U_1\!\!\oplus \!\!\beta_3U_2,X_3|Y_3)\label{eq:cov3S}\\
 &R_3\leq I(X_3;Y_3,\alpha_3U_1\oplus \beta_3U_2)\label{eq:cov3F}
\end{align}
,where $r_{\alpha,\beta}\triangleq \sum_{i\in[1,m]}\frac{k_i}{n}H(\alpha V_1\oplus \beta V_2),\forall \alpha,\beta\in \mathbb{F}_q$.
\end{Definition}
\begin{thm}
 A rate triple $(R_1,R_2,R_3)$ is achievable if it belongs to $cl(\mathcal{R}_{3\text{-IC}})$.
\end{thm}
\begin{IEEEproof}
We provide an outline of the proof. The coding scheme is similar to the one in \cite{IC}. Except that 1) decoder 2 also decodes a linear combination $\alpha_2 U_{1}+\beta_2 U_2$, 2) The underlying codes for $U_1$ and $U_2$ are QNLC's instead of nested coset codes, and 2) There is an outer code on $U_2$ which allows decoder $2$ to decode $U_2$ from $\alpha_2 U_{1}+\beta_2 U_2$. As a result the rate region is similar to the one in \cite{IC} except for a few changes. Bounds (\ref{eq:packS})-(\ref{eq:packF}) ensure the existence of jointly typical codewords at each encoder. These bounds are the same with the ones in \cite{IC}. Bounds  (\ref{eq:cov1S})-(\ref{eq:cov1F}) ensure errorless decoding at decoder $1$, they also remain the same. Inequalities (\ref{eq:cov2S})-(\ref{eq:cov2F}) correspond to the error events at decoder 2, these bounds are altered to ensure reconstruction of $\alpha_2U_1+\beta_2U_2$, also the rate $R_2$ is changed and the linear coding rate $T_2$ is multiplied by $\frac{I(U_2;\alpha_2U_1\oplus \alpha_2U_2)}{H(U_2)}$, which is due to the outer code. Lastly, (\ref{eq:cov3S})-(\ref{eq:cov3F}) are for the error events at decoder 3, which is also similar to the ones in \cite{IC}. 
\end{IEEEproof}

\begin{Remark}
 For ease of notation, we have dropped the time-sharing random variable $Q$. The scheme can be enhanced by adding the variable in the standard way. 
\end{Remark}
\begin{Remark}
 By taking $\alpha_2=0$ and $\beta_2=1$ and choosing the NQLC parameters so that the codes become a pair of NLC's we recover the bound in \cite{IC} as expected. 
\end{Remark}
\begin{Remark}
 Following the generalizations in \cite{IC}, this coding scheme can be enhanced by adding additional layers containing the public message codebooks corresponding to the HK strategy. \end{Remark}
\section{Conclusion}
\label{sec:conclude}
The problem of three user IC was considered. We showed that there is an inherent tradeoff in the general IC. The users can choose to communicate their messages by using optimal PtP strategies or cooperate with other users to facilitate their communication.  It was shown that the previously used coding structures are unable to optimize this tradeoff. New coding structures were proposed. It was shown through an example that these new structures give strict improvements. Using these new codebooks, an achievable region for the three user IC was derived which improves upon the previous known inner bounds for this problem.
\appendices
\section{Proof of Lemma \ref{Lem:eg1}}
\label{App:eg1}
 Assume the family $(n,\mathsf{M}_1,\mathsf{M}_2,\mathsf{M}_3, \mathbf{e},\mathbf{d}), n \in \mathbb{N}$ achieves the rate-triple. Let $M_i,i\in \{1,2,3\}$ be uniform random variables defined on sets $\mathsf{M}_i$. In the first step, we argue that the size of the set ${\mathbb{C}_1\oplus \mathbb{C}_2\oplus \mathbb{C}_3\oplus A_{\epsilon}^n(N_3)}$ is close to ${|\mathbb{C}_1\oplus \mathbb{C}_2\oplus A_{\epsilon}^n(N_3)||\mathbb{C}_3|}$. More precisely, we prove the following claim:

\begin{Claim}
For every $\epsilon\in \mathbb{R}^+$, there exists a sequence of numbers $\alpha_{n,\epsilon}\in \mathbb{R}^+, n\in\mathbb{N}$ such that the following inequality holds:
 
\begin{equation*}
 \frac{1}{n}\log{|\mathbb{C}_1\oplus \mathbb{C}_2\oplus \mathbb{C}_3\oplus A_{\epsilon}^n(N_3)|}\geq  \frac{1}{n}\log{(|\mathbb{C}_1\oplus \mathbb{C}_2\oplus A_{\epsilon}^n(N_3)||\mathbb{C}_3|)}-\alpha_{n,\epsilon},
 \end{equation*}
and $\alpha_{n,\epsilon}$ goes to 0 as $n\to \infty$ and $\epsilon \to 0$.
 \end{Claim}
 
\begin{IEEEproof}
 Intuitively, if the size of $\mathbb{C}_1\oplus \mathbb{C}_2\oplus \mathbb{C}_3\oplus A_{\epsilon}^n(N_3)$ is much smaller than $|\mathbb{C}_1\oplus \mathbb{C}_2\oplus A_{\epsilon}^n(N_3)||\mathbb{C}_3|$, that means there exists a large number of sets of vectors $\mathbf{c_1},\mathbf{c}_2,\mathbf{c}_3,\mathbf{n}_3$, with different $\mathbf{c}_3$'s for which the sum is equal. This causes a large error probability in decoder 3 since the decoder is unable to distinguish between these sets of vectors. More precisely, let $\mathbf{n_t}$ be a type on vectors in $A_{\epsilon}^n(N_3)$, and let $\mathbf{c}_3\in \mathbb{C}_3$, define $\mathsf{B}_{\mathbf{c}_3, \mathbf{n}_t}$ as follows, 
 \begin{equation*}
 \mathsf{B}_{\mathbf{c}_3,\mathbf{n}_t}=\{\mathbf{c}_1\oplus\mathbf{c}_2\oplus \mathbf{n}_3| \exists \mathbf{c}'_1,\mathbf{c}'_2,\mathbf{c}'_3, \mathbf{n}'_3\in \mathbb{C}_1\times\mathbb{C}_2\times\mathbb{C}_3\times \mathcal{P}_t,
  \text{ such that } \mathbf{c}'_1\oplus\mathbf{c}'_2\oplus\mathbf{c}'_3\oplus\mathbf{n}'_3=\mathbf{c}_1\oplus\mathbf{c}_2\oplus\mathbf{c}_3\oplus\mathbf{n}_3, \mathbf{c}'_3\neq\mathbf{c}_3\},
\end{equation*}
 where $\mathcal{P}_t$ is the set of all vectors $\mathbf{n}_3\in A_{\epsilon}^n(N_3)$ with type $\mathbf{n}_t$.
 That is $\mathsf{B}_{\mathbf{c}_3,\mathbf{n}_t}$ is the set of $(\mathbf{c}_1,\mathbf{c}_2,\mathbf{n}_3)$'s for which the decoder has non-zero error probability for decoding $\mathbf{c}_3$ or another codeword $\mathbf{c}'_3$.
  From set theory, we have the following:
 \begin{align}
 |\mathbb{C}_1\oplus \mathbb{C}_2\oplus \mathbb{C}_3\oplus A_{\epsilon}^n(N_3)|&\geq  |\mathbb{C}_1\oplus \mathbb{C}_2\oplus \mathbb{C}_3\oplus \mathcal{P}_t|\\
 &=  |\bigcup_{\mathbf{c}_3}\mathbb{C}_1\oplus \mathbb{C}_2\oplus \mathbf{c}_3\oplus \mathcal{P}_t|\\
 &\geq |\bigcup_{\mathbf{c}_3}\left(\mathbb{C}_1\oplus \mathbb{C}_2\oplus \mathbf{c}_3\oplus \mathcal{P}_t - \bigcup_{\mathbf{c}'_3\neq \mathbf{c}_3}   \mathbb{C}_1\oplus \mathbb{C}_2\oplus \mathbf{c}'_3\oplus \mathcal{P}_t  \right)|\\
 &=\sum_{\mathbf{c}_3} |\mathbb{C}_1\oplus \mathbb{C}_2\oplus \mathbf{c}_3\oplus \mathcal{P}_t - \bigcup_{\mathbf{c}'_3\neq \mathbf{c}_3}   \mathbb{C}_1\oplus \mathbb{C}_2\oplus \mathbf{c}'_3\oplus \mathcal{P}_t|\\
 &=\sum_{\mathbf{c}_3\in\mathbb{C}_3}(|\mathbb{C}_1\oplus \mathbb{C}_2\oplus \mathcal{P}_t|-|\mathsf{B}_{\mathbf{c}_3,\mathbf{n}_t}|)\\
 &=|\mathbb{C}_1\oplus \mathbb{C}_2\oplus \mathcal{P}_t||\mathbb{C}_3|-\sum_{\mathbf{c}_3\in\mathbb{C}_3}|\mathsf{B}_{\mathbf{c}_3,\mathbf{n}_t}|\\
 &=|\mathbb{C}_1\oplus \mathbb{C}_2\oplus \mathcal{P}_t||\mathbb{C}_3|-{|\mathbb{C}_3|}\sum_{\mathbf{c}_3\in\mathbb{C}_3}\frac{|\mathsf{B}_{\mathbf{c}_3,\mathbf{n}_t}|}{|\mathbb{C}_3|}\\
 &=\left(|\mathbb{C}_1\oplus \mathbb{C}_2\oplus \mathcal{P}_t|-E(|\mathsf{B}_{\mathbf{c}_3,\mathbf{n}_t}|)\right){|\mathbb{C}_3|} \\
 &=\left(|\mathbb{C}_1\oplus \mathbb{C}_2\oplus \mathcal{P}_t|(1-\frac{E(|\mathsf{B}_{\mathbf{c}_3,\mathbf{n}_t}|)}{|\mathbb{C}_1\oplus \mathbb{C}_2\oplus \mathcal{P}_t|})\right){|\mathbb{C}_3|}\label{lastneq}
.
\end{align}
On the other hand, as $n\to \infty$, the error probability at decoder $3$ goes to 0. This means that $P\left(\mathbf{d}_3(\mathbf{c}_1\oplus \mathbf{c}_2\oplus \mathbf{e}(M_3) \oplus \mathbf{n}_3)\neq M_3      \right)$ goes to 0. Consequently, there exists a family of types type $\mathbf{n}_t$ such that $P\left(\mathbf{d}_3(\mathbf{c}_1\oplus \mathbf{c}_2\oplus \mathbf{e}(M_3) \oplus \mathbf{n}_3)\neq M_3   |\mathbf{n}_t   \right)$ goes to 0. There exists a sequence $\delta_n$ which approaches 0 at the limit such that:
\begin{align*}
 \delta_n &\geq P\left(\mathbf{d}_3(\mathbf{c}_1\oplus \mathbf{c}_2\oplus \mathbf{e}(M_3) \oplus \mathbf{n}_3)\neq M_3 |\mathbf{n}_t     \right)\\
 &{\geq} \frac{1}{2}P\left(\mathbf{c}_1\oplus\mathbf{c}_2\oplus \mathbf{n}_3\in \mathsf{B}_{\mathbf{e}(M_3),\mathbf{n}_t}|\mathbf{n}_t \right) \\
 &{\geq} \frac{1}{2}\sum_{\mathbf{c}_3\in \mathbb{C}_3}\frac{|\mathsf{B}_{\mathbf{c}_3,\mathbf{n}_t}|}{|\mathbb{C}_3||\mathbb{C}_1\oplus \mathbb{C}_2\oplus \mathcal{P}_t|}\\
&=\frac{1}{2}\frac{E(|\mathsf{B}_{\mathbf{e}_3(M_3),\mathbf{n}_t}|)}{|\mathbb{C}_1\oplus \mathbb{C}_2\oplus \mathcal{P}_t|} 
\end{align*}

Inserting this last inequality in (\ref{lastneq}) we have, 
\begin{align*}
 &|\mathbb{C}_1\oplus \mathbb{C}_2\oplus \mathcal{P}_t||\mathbb{C}_3|-\sum_{\mathbf{c}_3\in\mathbb{C}_3}|\mathsf{B}_{\mathbf{c}_3}|\\
 &\geq \left(|\mathbb{C}_1\oplus \mathbb{C}_2\oplus \mathcal{P}_t|(1-2\delta_n) \right){|\mathbb{C}_3|}
\end{align*}

\end{IEEEproof}

 Observe that $\mathcal{P}_t$ and $A_{\epsilon}^n(N_3)$ have the same exponential rate. Note that $|\mathbb{C}_1\oplus \mathbb{C}_2\oplus A_{\epsilon}^n(N_3)|\geq|\mathbb{C}_1\oplus A_{\epsilon}^n(N_3)|$ and since decoder 1 can decode $X_1$ with probability of error approaching 0, we can use the same argument to show the following:
 
\begin{align}
  &\frac{1}{n}\log{|\mathbb{C}_1\!\!\oplus A_{\epsilon}^n(N_3)|}\!\!\to\!\! \frac{1}{n}\log{|\mathbb{C}_1||A_{\epsilon}^n(N_3)|}\to \!\log{q}\!-\!\!H(N_1)\oplus H(N_3)\nonumber \\
  &\Rightarrow\frac{1}{n}\log{|\mathbb{C}_1\oplus \mathbb{C}_2\oplus \mathbb{C}_3\oplus A_{\epsilon}^n(N_3)|}\geq 
  \\&\qquad\qquad\left(\log{q}-H(N_1)+H(N_3)\right)+\left(H(N_1)-H(N_3)\right)=\log{q}.\nonumber
\end{align}
But we know that $\frac{1}{n}\log{|\mathbb{C}_1\oplus \mathbb{C}_2\oplus \mathbb{C}_3\oplus A_{\epsilon}^n(N_3)|}\leq\log{q}$. So, we should have equality at all of the inequalities. Hence, $\frac{1}{n}\log_q{|\mathbb{C}_1\oplus \mathbb{C}_2\oplus A_{\epsilon}^n(N_3)|}\to \frac{1}{n}\log_q{|\mathbb{C}_1\oplus A_{\epsilon}^n(N_3)|}$. In order to have  $|\mathbb{C}_1\oplus \mathbb{C}_2|$ close to $|\mathbb{C}_1|$, we must have the properties stated in the lemma. 
\section{Proof of Lemma \ref{Lem:newrate}}
\label{App:newrate}
We provide a coding scheme based on NQLC's which achieves the rate vector. Consider two ternary random variables $V_1$ and $V_2$ such that $H(V_1\oplus V_2)>H(2V_1\oplus V_2)$. We will show the achievability of the following rate-triple:
 \begin{align*}
& R_1=\log{q}-H(N_1\oplus N_2\oplus N_3)\\
& R_2=(\frac{H(V_1\oplus V_2)}{H(V_1)}-1)(\log{q}-H(N_1\oplus N_2\oplus N_3))\\
& R_3=\log{q}-H(N_3)-\frac{H(2V_1\oplus V_2)}{H(V_1)}\left(\log{q}-H(N_1\oplus N_2\oplus N_3)\right).
\end{align*}
Note that in this case $R_2+R_3=H(N_1\oplus N_2\oplus N_3)-H(N_3)+\frac{(H(V_1\oplus V_2)-H(2V_1\oplus V_2))}{H(V_1)}\left(\log{q}-H(N_1\oplus N_2\oplus N_3)\right)$. Choose random variable $V_3$ such that $R_3=H(V_3\oplus N_3)-H(N_3)$.
 \\\textbf{Codebook Generation:} Construct a family of pairs NQLC's with length $n$ and parameters $m=1$, $k_1=\frac{(\log{q}-H(N_1\oplus N_2\oplus N_3))}{H(V_1)}n$, $U_1=V_1$, and $U'_1=V_2$ by choosing the dither $\mathbf{b}$ and generator matrix $G_1$ randomly and uniformly on $\mathbb{F}_q$. For a fixed $n\in \mathbb{N}$, Let $\mathcal{C}^n_1$ and $\mathcal{C}^n_2$ be the corresponding pair of NQLC's. Let $\Phi_i=2^{nR_i}$ for $i\in \{1,2\}$. Choose $\Phi_i$ of the codewords in $\mathcal{C}_i^n$ randomly and uniformly, and index these sequences using the indices $[1,\Phi_i]$. Also, generate an unstructured codebook $\mathcal{C}_3$ randomly and uniformly with rate $R_3$ based on the single-letter distribution $P_{V_3}$. Index $\mathcal{C}_3$ by $[1,2^{nR_3}]$.
 \\\textbf{Encoding:} Upon receiving message index $M_i$ encoder $i$ sends the sequence in $\mathcal{C}_1$ which is indexed $M_i$ for $i\in\{1,2\}$. Let the codewords sent by encoder $i, i\in \{1,2\}$ be denoted by $\mathbf{v}_iG_1\oplus \mathbf{b}_i$. 
 Encoder 3 sends the codeword in $\mathcal{C}_3$ indexed by $M_3$. Let the codeword sent by the third decoder be denoted by $\mathbf{c}_3$.
 \\\textbf{Decoding:} Decoder 1 receives $X^n_1\oplus N^n_1\oplus N^n_2\oplus N^n_3$. Using typicality decoding, the decoder can decode the message as long as $\frac{k_1}{n}H(V_1)\leq \log{q}-H(N_1\oplus N_2\oplus N_3)$.  Decoder 2 receives $X^n_1\oplus X^n_2\oplus N^n_2\oplus N^n_3=(\mathbf{v}_1\oplus \mathbf{v}_2)G_1\oplus \mathbf{b}_1\oplus \mathbf{b}_2\oplus N^n_2\oplus N^n_3$. It can decode $\mathbf{v}_1,\mathbf{v}_2$ jointly as long as 1) $ \frac{k_1}{n}H(V_1\oplus V_2)<\log{q}-H(N_1\oplus N_3)$, and 2) $R_1+R_2\leq \frac{k_1}{n}H(V_1\oplus V_2)$. The first condition ensures that $\mathbf{v}_1\oplus \mathbf{v}_2$ can be recovered with probability of error going to 0 as $n\to \infty$. After recovering $\mathbf{v}_1\oplus \mathbf{v}_2$, the decoder needs to jointly decode $\mathbf{v}_1,\mathbf{v}_2$ (for reasons explained in Lemma \ref{Lem:lem2}). This is a noiseless additive MAC problem and condition 2 ensures errorless decoding. Note that in condition 2, the coefficient $\frac{k_1}{n}$ is present since $\mathbf{v}_1$ is of length $k_1$. Also, The term $H(V_1\oplus V_2)$ is the capacity of the MAC channel. Decoder 3 receives $X^n_1\oplus X^n_2\oplus X^n_3\oplus N^n_3=(2\mathbf{v}_1\oplus \mathbf{v}_2)G_1\oplus \mathbf{b}_1\oplus \mathbf{b}_2\oplus \mathbf{c}_3^n\oplus N^n_3$. The decoder can recover $2\mathbf{v}_1\oplus \mathbf{v}_2$ as long as $\frac{k_1}{n}H(2V_1\oplus V_2)<\log{q}-H(X_3\oplus N_3)$. Then, the decoder subtracts $2X_1^n\oplus X_2^n$ to get $X^n_3\oplus N^n_3$. It can decode $X_3$ as long as $R_3\leq H(V_3\oplus N_3)-H(N_3)$.
 It is straightforward to check the rate given at the beginning satisfy all of these bounds.

\end{document}